\documentstyle{laa}
\begin{document}
\thesaurus{13. 
    (02.07.1,02.09.1,05.03.1,11.05.2)}
\title{On the stability of motion of $N$-body systems:
the effect of the variation of particle number, softening and rotation}
\author{A.A. El-Zant$^{1,2}$}
\institute{Astronomy Centre, University of Sussex,
Brighton BN1 9QH, UK\and Physics Department, Technion --- Israel Institute of Technology, Haifa 32000, Israel}
\date{Received .........; accepted.........}
\maketitle

\begin{abstract}   
           
Using the Ricci and scalar curvatures of the configuration manifold of 
gravitational $N$-body systems,
 we study the exponential instability in their trajectories. 
It is found that the exponentiation time-scale for isotropic 
Plummer spheres varies very little with particle number if the softening
is small.  Large softening on the other hand
has a marked effect and, if large enough, can cause the curvatures to become
positive. This last result confirms the  previous observations  for 
self gravitating sheets and suggests that the {\em qualitative} behaviour 
 of large-$N$ and continuum systems may be different, and that their equivalence
is only obtained in the limit of infinite $N$ and finite softening. 
It is also found that the presence of a large fraction of the kinetic energy in
rotational motion increases the exponentiation time-scales significantly --- an
effect that should be expected given the regular nature of nearly circular motion.
  In the light of the results of
this and of previous studies, it is   suggested that the exponential instability
 may arise from low order resonances between the period of the variation of
the gravitational field due to distant encounters and the orbital period of a test
particle. For periods long compared to the exponentiation time
but short compared to the diffusion time-scales of the action variables, the  standard
picture of collisionless dynamics may  be valid in an averaged sense --- nevertheless
this time interval need not coincide with that predicted by standard relaxation theory. 
Instead it is suggested that, at least for systems with well defined final states, the
relaxation time should scale as $\sim N^{1/2}$.

\end{abstract}

\section{Introduction}
\label{intro}
It is well documented in numerical simulations that $N$-body gravitational systems
display an exponential instability with respect to small changes in the initial conditions 
(e.g., Miller 1964; Goodman et al. 1993; Kandrup et al. 1994). This 
instability not only
appears when the linearised dynamics is studied but also when  the full nonlinear
evolution of two originally similar systems is considered. It does not appear therefore
to be simply a product of linearisation of the equations of motion --- a linearisation 
instability. It is possible however (El-Zant 1996a) that the exponential instability
observed  in short time calculations of
the divergence of temporal states may result from phase mixing.
However there is another way in which the instability manifests itself, namely
through the predominantly negative two dimensional curvature of the configuration 
 space of $N$-body systems (Gurzadyan \& Savvidy 1984,1986; Kandrup 1990a, 1990b).  
This property arises from the qualitative phase space structure
of a system and cannot be explained away so easily.
It necessarily implies instability {\em normal} and not just along the phase 
space\footnote{The term 
phase space in this paper will refer to the 6 $N$ dimensional phase space 
unless otherwise indicated.} trajectory of
a system (a property known in dynamical literature as transversality: 
e.g., Ruelle 1989), which implies
{\em qualitatively} different behaviour from  that of regular systems. 
Since, if real, such an effect could have far reaching implications on the
evolution of gravitational systems, it is natural to enquire as to how the predicted
instability correlates
with various parameters of $N$-body systems in an attempt to uncover its origin and
implications.
Since not all two dimensional curvatures  are always negative during the evolution
of these systems, and in any case, these depend on the full Riemann tensor
(difficult to calculate for all but the smallest systems) an averaged chaos indicator
still relying on the geometric approach has to be used. 

Such a method, involving  the  Ricci (or mean) curvature, 
was devised by Gurzadyan \& Kocharyan (1987) and
 was applied to numerically integrated $N$-body systems 
by El-Zant (1996a) and  El-Zant \& Gurzadyan (1997).
In these papers 
it was shown that the stability of motion  of $N$-body gravitational
systems as described by the aforementioned method correlated strongly with various
parameters of a gravitational system. For although the exponential instability (as evidenced
by the negative Ricci curvature) 
 exists for most initial conditions, it was  observed for example 
that this instability was more pronounced when clear macroscopic 
(e.g., collective plasma type) instabilities were 
present  and in situations where one expects a faster evolution rate
 (e.g., in the presence of central concentration which accellerates the
gravothermal evolution of gravitational systems towards still more concentrated
states with higher thermodynamic entropy: Saslaw 1985).
 On the other hand, the predicted  instability  was much weaker for regular
systems (e.g. for ones in uniform rotation before macroscopic 
instability starts to have 
a significant effect).  
Moreover, it was found (El-Zant 1996a) that the  negativity
of the Ricci curvature is not a result of contributions due to close encounters but
in fact in spite of them --- the curvature became always negative only when the
fluctuations due to the closest encounters were removed. In fact, for a system 
in statistical equilibrium, the Ricci curvature is constant 
(up to random fluctuations and neglecting close encounters)
and negative and mainly determined by the first derivatives of the potential. 
This suggests that the instability arising from this property is unlikely to be the
result of the linearisation of a system containing fluctuating forces.
In other words, it is likely to be the manifestation of chaotic behaviour characteristic
of generic $N$-body gravitational systems. 
 It is often assumed that if this is the case then the effect of the instability
mainly concerns the accuracy of numerical simulations of $N$-body systems
(e.g., Goodman et al. 1993; Miller 1994).  This
problem is also mentioned in many papers dealing with numerical simulation 
techniques (e.g.,
Aarseth 1996; Barnes \& Hut 1989). Chaos however does not only affect numerical simulations, it
usually comes with other deep implications. The most important is that it leads to diffusion 
in the action variables which determine the physical state of a system.
This diffusion time-scale need not be determined by the classical two body
theory --- even if it is long compared to the exponentiation time --- since
this theory assumes an integrable system which remains so under perturbations
due to discreteness. On the other hand, systems with  negative curvature (whether constant or not) usually have strong statistical properties, very different from near 
integrable systems (e.g., Pesin 1989).
It is therefore very plausible that there may be, contrary to what is
often assumed, a physical meaning to the exponential instability. 
This meaning is however not yet completely clear. 

In this paper we continue our investigation of the stability of motion of $N$-body
systems using the Ricci and scalar curvatures  of the configuration
manifold of generalised coordinates (the method is briefly described in Section~\ref{method}).  
We will be interested here in the case of spherical systems. The main 
questions we would like to ask are as follows. First, it has been observed 
(both in direct simulations and by using the Ricci curvature method) that 
the exponentiation time-scales are rather small (usually a fraction of a dynamical
time). If this time-scale does not increase with $N$ then one must conclude 
that it cannot be directly related to the diffusion of the action variables
in a system in virial equilibrium --- and therefore 
any physical meaning  will have
to be more subtle. In Section~\ref{sphex:secn}  we examine the variation of that
time-scale for $N$ up to 15 000. For  flat sheet like systems it was found (El-Zant 1996a) that significant
softening destroyed the negative Ricci curvature of the configuration space, meaning that
the structure of the phase-space should be different and  that the
motion should be more regular. If this is the case, then the exponential instability 
must arise from discreteness effects and there has to  be a discontinuity in the 
transition between the phase space structure of large (but finite) $N$-body gravitational
systems and their continuum counterparts.
In Section~\ref{soft} we study the effect of softening in spherical Plummer models and
its possible implications in some detail.
Another problem is concerned with the fact that many galaxies
are observed to be in fairly regular states where most of the kinetic energy is in the
form of ordered rotational motion. Again, a small exponentiation time-scale 
(if directly interpreted) is incompatible with such motion --- especially if
the divergence takes place from most initial conditions, in which case there
is no room for ``stable chaos'' or trapping between KAM tori (as for example suggested by
 Gouda et al. 1994). 
It is therefore
important  to see if the presence of large amount of rotational motion does influence the
relaxation time-scales significantly. This effect is studied in Section~\ref{sphex:rotsy}.
In the final section we summarise the results and  describe  
a possible interpretation of
the origin and effect of the exponential instability of $N$-body systems and how this can be tested.
 
\section{Method}
\label{method}
The study of the stability properties of  gravitational $N$-body systems using
the Ricci curvature was initiated by Gurzadyan \& Kocharyan (1987).
Details of how the Ricci curvature can be used  to study the stability of $N$-body 
trajectories are given in El-Zant (1996a). We just mention here a few points that
are essential to the interpretation of the results of the following sections.

Through the Lagrangian formulation of dynamics, it is possible to reduce the study
of the stability of motion of $N$-body systems to that of the geometry of the
corresponding Lagrangian
manifold. When this is done, the Gaussian curvatures on any two dimensional directions
normal to the motion 
 are found to be mostly (but not always)
negative (e.g., Gurzadyan \& Savvidy 1986; Kandrup 1990a). Since for large $N$ the
probability any of the two 
dimensional curvatures  being positive becomes exceedingly small,
it is possible to replace
the full set of two dimensional curvatures by their average over the $3N -1$ directions
normal to the geodesic velocity vector  ${\bf u}$ . This quantity also happens to
coincide with the Ricci curvature of the manifold. In Cartesian coordinates
in the enveloping $3N$ configuration space this is given by
\begin{equation}
r_{u}=3A\frac{\left(W_{i}u^{i}\right)^{2}}{W^{2}}
-2A\frac{W_{ij}u^{i}u^{j}}{W}
-\left(A-\frac{1}{2} \right)\frac{\parallel \nabla W \parallel^{2}}{W^{3}}
-\frac{\nabla^{2}W}{2W^{2}} 
\label{eq:ru}
\end{equation}
with $A=\frac{3N-2}{4}$.
Here $W$ is the kinetic energy
calculated as a function of the potential energy while  $\nabla W= \sum W_{i}$ 
and $\nabla^{2} W= \sum W_{ii}$.
For a $N$-body system,  the implied summation would be over $i,j=1,3N$.  
If we label 
by $a$, $b$ and $c$ the particle numbers (which run from 1 to $N$)
and by $k$ and $l$ the three Cartesian coordinates of a particle, then
for particles with same mass $m$ and with $G=1$ we have
\begin{equation}
W_{i}=\frac{\partial W}{\partial q_{i}}=\frac{1}{\sqrt{m}}\frac{\partial W}{\partial
r^{k}_{a}}=- m^{3/2} \sum_{c \neq a} \frac{r^{k}_{ac}}{r^{3}_{ac}},
\end{equation}
\begin{equation}
W_{ij}=\frac{\partial^{2} W}{\partial q_{i} \partial q_{j}}=
\frac{1}{m} \frac{\partial^{2} W}{\partial r^{k}_{b} \partial r^{l}_{a}}=
m \left[\frac{\delta_{kl}}{r^{3}_{ab}}- \frac{3r^{k}_{ab}
r^{l}_{ab}}{r^{5}_{ab}} \right],     \label{eq:ruw}
\label{eq:rut1}
\end{equation}
if $a \neq b$ and
\begin{equation}
W_{ij}=\frac{\partial^{2} W}{\partial q_{i}  \partial q_{j}}=
\frac{1}{m} \frac{\partial^{2} W}{\partial r^{k}_{b} \partial r^{l}_{a}}=
- m \sum_{c \neq a} \left[\frac{\delta_{kl}}{r^{3}_{ac}}- \frac{3r^{k}_{ac}
r^{l}_{ac}}{r^{5}_{ac}} \right],
\label{eq:rut2}
\end{equation}
if $a = b$. In these equations
\begin{equation}
r^{2}_{ab}=(r^{1}_{ab})^{2}+(r^{2}_{ab})^{2}+(r^{3}_{ab})^{2} + b^{2},
\end{equation}
where $b$ is a softening parameter and \hspace{0.1in}$r^{k}_{ab}=r^{k}_a -r^{k}_b$. 

A negative Ricci curvature can be interpreted to imply that a system will,
in general, display exponential instability to random perturbations (El-Zant 1996b).
The associated exponentiation time-scale is then given by
\begin{equation}
\tau_{e} \sim \left( \frac{3N}{-2 \bar{r}_{\bf u}} \right)^{1/2} \times \frac{1}{\bar{W}},     
\label{dyn}
\end{equation}
where a bar denotes a phase space average (can be either static or dynamic).

When the two dimensional curvatures are averaged over all geodesics
that pass by a given point as well as  over the  directions normal
to the velocity vector at that point, one obtains the curvature scalar
\begin{equation}
R=\sum_{\bf u,n} k_{\bf u,n}= \sum_{\bf u} r_{\bf u}.
\label{eq:scalar}
\end{equation}
This quantity turns out to be always negative for $N>2$. This means that a $N$-body system 
will, in general, display exponential instability when randomly perturbed
and for random distribution  in velocity space. 
An associated 
instability time-scale can also be inferred: One replaces $r_{\bf u}$ in
Eq.~(\ref{dyn}) by $R$ and $3N$ by $(3N)^{2}$. Clearly the time-scale derived
from the scalar curvature can be equal to that inferred from
the Ricci curvature only if no ordered motion is present. We will see that
this is indeed the case. 

It is the exponentiation time-scales that will
 be of interest in the coming sections. We will
be averaging them either over (pseudo)random realizations of $N$-body sytems or 
over time-averages for numerically integrated systems. In the latter case, we
use the Aarseth (1996) NBODY2 code which is a direct summation code applying
the Ahmad \& Cohen (1973) neighbour scheme and individual time-steps for each
of the particles in the simulation. These refinements speed up the integration
considerably, while essentially maintaining the accuracy and simplicity of direct
summation codes. Since we are still in the exploratory
stages of applying and testing geometric methods  to gravitational systems,
it is prudent to study the behaviour of the curvatures when calculated in 
an accurate and straightforward manner before integrating it into large-$N$ codes.
The parameters of the code are fixed at the same
values used in El-Zant \& Gurzadyan (1997).

Finally, it is important to note here that the negativity of the curvatures is 
only one mechanism by which chaos can occur according to the geometric formulation
of dynamics. Another mechanism is provided by parametric instability 
as pointed out by Pettini (1993) and  Cerruti-Sola \& Pettini (1995). However,  when $N$ is large
and the system considered is  near virial equilibrium, the mean potential energy does not
undergo significant fluctuations. In that case the second and third terms of  
 Eq. (26) of the latter paper, which depend on the time derivatives of 
the potential energy, are 
then very small. This leaves negative curvature as the only effective mechanism for instability.

\section{Behaviour of the curvatures  as particle numbers change}
\label{sphex:secn}

In this section we look at the effect of varying the particle number on the
exponentiation time-scale of isotropic equilibrium Plummer models prepared 
using the method of Aarseth et al. (1974) and scaled to the units of Heggie \& Mathieu (1986) 
by keeping the total mass and the gravitational constant equal
to unity and the total energy fixed at $E=-0.25$. In this case the mean
 crossing time $\tau_{c}$ is equal to $2 \sqrt{2}$  time units.

Direct summation routines are not well suited for integration of large $N$-body systems
(except if used in conjunction with special hardware like GRAPE architecture
which was not available to the author) and NBODY2 becomes very slow for particle numbers 
exceeding a few thousand as to prevent systematic examination of the dynamics for greater
particle numbers. Fortunately however, the fact that (give or take random fluctuations) 
the curvatures are constant for systems in statistical dynamical equilibrium 
(Eq.~(\ref{eq:ru})) means that,
for such systems, it is possible to get an estimate of these quantities by simply 
calculating them for different (pseudo)random realizations of the same one particle
phase space distribution. This in turn is sufficient to give us an idea of the values
of the curvatures for fairly large spherical $N$-body systems in statistical dynamical
equilibrium. 

For systems consisting of up to $N=1400$,
we integrated the full equations of motion and calculated the Ricci and scalar curvatures
along the motion. This is done at intervals of a hundred starting for $N=100$. 
The  behaviour of the Ricci curvature time series during the evolution
of such systems is described in El-Zant \& Gurzadyan (1997). 
As in El-Zant \& Gurzadyan (1997)
we remove the contributions due to the closest encounters by introducing a short range 
cutoff in the calculation of the potential energy and its derivatives. 
We choose a distance (including the softening) of 0.05 for this
(i.e., $5 \%$ the virial radius). This is typically about a third 
of the minimum radius of the neighbour spheres as calculated by the NBODY2 algorithm.
For systems
consisting of up to $N=15000$ particles we  calculate  the Ricci and scalar curvatures for ten different realizations of the  same equilibrium distribution 
 and take the
average. This is done starting at $N=2000$ at intervals of a thousand.
To compare the results of the two approaches we  take the average over the first ten
outputs of the low $N$ runs. This corresponds to one crossing time. Over this time interval
the structure of the spheres are not expected to have changed much and, if indeed the 
motion is chaotic over small time-scales, these numbers also correspond to  pseudorandom
realisations of the same phase space distribution. 
The softening of the potential (which
also follows the  Plummer law) was taken as $b=2/N$ which scales  like the ratio 
of minimum to maximum impact parameters of standard relaxation theory 
(Binney \& Tremaine 1987; Farouki \& Salpeter 1994; Gierz \& Heggie 1994). 
This is calculated as to exclude 
encounters (assumed independent)  which lead to deflections beyond a certain
maximal value.

\begin{table}
\begin{center}
\begin{tabular}{c|c|c|c|c}
\multicolumn{5}{c}{ }\\
\hline
$N$        & $\tau_{er}$      & $\epsilon_{r}$       & $\tau_{es}$ & $ \epsilon_{s}$\\ 
\hline
100        & 0.28             & 0.78                 & 0.21        & 0.06\\
\hline
200        & 0.34             & 0.46                 & 0.23        & 0.04\\
\hline
300        & 0.32             & 0.13                 & 0.27        & 0.01\\
\hline                 
400        & 0.28             & 0.21                 & 0.29        & 0.01\\
\hline
500        & 0.31             & 0.17                 & 0.29        & 0.01\\
\hline
600        & 0.28             & 0.07                 & 0.29        & 0.01\\
\hline   
700        & 0.31             & 0.12                 & 0.29        & 0.01\\
\hline   
800        & 0.31             & 0.09                 & 0.30        & 0.01\\
\hline                 
900        & 0.29             & 0.07                 & 0.30        & 0.01\\
\hline
1000       & 0.31             & 0.09                 & 0.30        & 0.01\\
\hline
1100       & 0.30             & 0.07                 & 0.31        & 0.01\\
\hline   
1200       & 0.31             & 0.05                 & 0.31        & 0.01\\
\hline   
1300       & 0.32             & 0.05                 & 0.31        & 0.01\\
\hline                 
1400       & 0.31             & 0.05                 & 0.31        & 0.01\\
\hline
2000       & 0.29             & 0.00                 & 0.31        & 0.00\\
\hline
3000       & 0.31             & 0.00                 & 0.31        & 0.00\\
\hline   
4000       & 0.30             & 0.00                 & 0.31        & 0.00\\
\hline     
5000       & 0.31             & 0.00                 & 0.31        & 0.00\\
\hline
10000      & 0.31             & 0.00                 & 0.31        & 0.00\\
\hline
\hspace{0.1in} 15000    \hspace{0.1in}   &\hspace{0.1in}  0.31  \hspace{0.1in}           
&\hspace{0.1in}  0.00  \hspace{0.1in}    &\hspace{0.1in}  0.31 \hspace{0.1in}       
&\hspace{0.1in} 0.00 \hspace{0.1in} 
\end{tabular}
\end{center}
\caption{\label{sphex:tabN} Variation  with particle numbers
of the exponentiation time-scales $\tau_{er}$ and
$\tau_{es}$ calculated from the values of the Ricci and scalar curvatures. In the third 
and and fifth columns 
the associated errors estimated with the aid of Eq.~(\ref{sphex:ercur})
 are given}
\end{table}

Except in the case of $N=100$ when fluctuations are fairly large, causing  
it to be occasionally positive, it is found that the Ricci curvature
(calculated while excluding the contribution from neighbouring members as described above)
is {\em always} negative.  It is therefore easy to extract an average 
exponential divergence time-scale through the method described in the previous section.
The results are shown in Table~\ref{sphex:tabN}. In the first
column are the particle numbers.   
In the second column and fourth columns are the exponentiation time-scales (in crossing times) 
averaged over ten different values of the Ricci or Scalar curvatures as described above,
while  the third and fifth columns  contain  estimates of the RMS relative  dispersion in the
ten  calculated exponentiation time-scales. This is obtained from
\begin{equation}
\epsilon=\frac{ \sqrt{ \frac{\sum (\tau^{i}_{e}  - \bar{\tau}_{e})^{2}}{10} } }{ \bar{\tau}_{e} } ,
\label{sphex:ercur}
\end{equation}
where $\bar{\tau}_{e}$ denotes the mean of the ten values.

Two things are immediately clear from these results. 
The first is that the exponentiation time-scale is quite short --- being a fraction of
the crossing time.
The second is that there appears
to be  no increase in the exponentiation time-scale as $N$ increases. This is true of both the 
estimates using the Ricci and the scalar curvatures. 
Although the exponentiation time-scales predicted  by the latter quantity slightly increase 
with $N$ when this number is limited to a few hundred particles, they quickly converge 
to the same value predicted by the Ricci curvature for larger $N$. 

Perplexing as these results may seem, their numerical explanation 
is actually relatively straightforward. First,
since the Ricci curvature is mainly determined by the bulk properties of the 
system and not by fluctuations due to nearest neighbours, one expects its average over $N$
to remain constant with (large) increasing $N$, provided that these global properties (described
by the one particle distribution function of the system) 
remain unchanged. This means that the exponentiation time-scale,
which depends on that average (cf. Eq.~(\ref{dyn})), remains constant for large enough $N$. 
The order of magnitude of the exponentiation time-scale can be understood as follows.
 The  last term on the
right hand side of Eq.~(\ref{eq:ru}) is very small when the softening is --- the
  situation we are interested in here. 
The first term is also small if the forces on the particles 
are not aligned with their velocities (a very improbable situation in an
equilibrium configuration). 
We are therefore left with the  second and third terms in the expression
for $r_{\bf u}$. For systems near virial equilibrium, 
and in the absence of ordered motion,
the second term is dominated by contributions from close encounters.
It is highly fluctuating and averages to zero when the softening is small
(this statement however does not appear to hold
when ordered motion is present as we shall see in Section~\ref{sphex:rotsy} below). 
However, although its time averaged contribution is very small, it can have large positive values
and, for small $N$, it occasionally causes the Ricci curvature to be positive if no short range
cutoff is introduced. When such a cutoff is introduced however, this term
is always small compared to the third. It also becomes small compared to the third term when the
particle numbers are large, thus supporting the 
 predictions of Gurzadyan \& Savvidy (1984,1986) and Kandrup (1990a, 1990b) that as $N$ increases the curvature is more likely to become negative.
This can be seen from Table~\ref{sphex:tabH} where the values
of the two terms are shown for single (pseudo)random realizations of Plummer spheres consisting
of up to 45 000 unsoftened particles (and no short range cutoff in the calculation of the 
potential and its derivatives).

\begin{table}
\begin{center}
\begin{tabular}{c|c|c|c}
\multicolumn{4}{c}{ }\\
\hline
$N$ & $2A\frac{W_{ij}u^{i}u^{j}}{W}$  
& - $ \left(A-\frac{1}{2} \right)\frac{\parallel \nabla W \parallel^{2}}{W^{3}} $ & $r_{\bf u}$\\ 
\hline
20 000        & $-1.5 \times 10^{5}$   & $ -3.3 \times 10^{5}$    & $ -4.9 \times 10^{5}$ \\              
\hline
25 000        & $-1.5 \times 10^{5}$   & $- 3.9 \times 10^{5}$    & $-5.6 \times 10^{5}$\\                
\hline                 
30 000        & $-9.4 \times 10^{4}$   & $ - 4.6 \times 10^{5}$   & $-5.6 \times  10^{5}$ \\                
\hline
35 000        & $ 1.7 \times 10^{5}$   & $ -5.3 \times 10^{5} $   & $-3.5 \times 10^{5} $\\                
\hline
40 000        & $ 3.0 \times 10^{4} $  & $ -6.0 \times 10^{5} $    & $-5.7 \times 10^{5}$ \\                 
\hline   
45 000        & $ 2.4 \times 10^{5} $     & $-7.4 \times  10^{5} $    & $ -4.9 \times 10^{5}$                                                          
\end{tabular}
\end{center}
\caption{\label{sphex:tabH} Magnitudes of the second
 and third term in Eq.~(\ref{eq:ru}) 
and of the resulting Ricci curvature for large $N$. 
Values are obtained from a single
pseudorandom realizations of the Plummer models}
\end{table}

The above explains why the exponentiation time-scales described by the Ricci and Scalar 
curvatures are similar since it is the third term that appears in the formula of the 
scalar curvature (e.g., Gurzadyan \& Savvidy 1986). 
A simple argument, due to Kandrup (1989), shows why these time-scales should not vary much 
with particle number.  Adopted for use in conjunction with the Ricci curvature, 
it goes as 
follows. For large $N$ the Ricci curvature dominated by the last term becomes
\begin{equation}
r_{\bf u}\sim \frac{3N}{4} N \times \left( \frac{\bar{F}}{W} \right)^{2},
\end{equation} 
where $\bar{F}$ denotes the average total RMS force per unit mass acting on a test particle. Using 
Eq.~(\ref{dyn}), this implies an exponentiation time-scale of $\tau_{e} \sim \bar{v}/\bar{F}$, where
$\bar{v}$ is the RMS speed. This time-scale is of course of the order of a dynamical time.   

\section{Behaviour of the curvatures as the softening parameter is increased}
\label{soft}

In the case of $N=1000$, we have calculated the Ricci and scalar curvatures for a range of softening
radii starting from $10^{-3}$ units (i.e, one thousandth of the virial radius)  to 
$4 \times 10^{-2}$ units. In Fig~\ref{softcur} we plot the Ricci curvature as a function of the
crossing time for
some representative values of the softening parameter. These plots show that  as this 
parameter increases so does the Ricci curvature, eventually it becomes positive
just as in the case of the flat systems (El-Zant 1996a).

\begin{table}
\begin{center}
\begin{tabular}{c|c|c|c|c}
\multicolumn{5}{c}{ }\\
\hline
$b$  & $\tau_{er}$  &  $\epsilon_{r}$ & $\tau_{es}$ &  $\epsilon_{s}$\\ 
\hline
0.001        & 0.30             & 0.09                 & 0.30        & 0.02\\
\hline
0.002        & 0.30             & 0.07                 & 0.30        & 0.02\\
\hline
0.004        & 0.31             & 0.08                 & 0.30        & 0.02\\
\hline                 
0.006        & 0.31             & 0.09                 & 0.30        & 0.01\\
\hline
0.008        & 0.31             & 0.09                 & 0.30        & 0.02\\
\hline
0.010        & 0.32             & 0.08                 & 0.30        & 0.02\\
\hline   
0.012        & 0.33             & 0.09                 & 0.32        & 0.02\\
\hline   
0.014        & 0.34             & 0.08                 & 0.33        & 0.02\\
\hline                 
0.016        & 0.35             & 0.10                 & 0.34        & 0.02\\
\hline
0.018        & 0.37             & 0.11                 & 0.35        & 0.02\\
\hline
0.020        & 0.40             & 0.13                 & 0.37        & 0.02\\
\hline   
0.022        & 0.42             & 0.12                 & 0.39        & 0.02\\
\hline   
0.024        & 0.47             & 0.17                 & 0.41        & 0.03\\
\hline                 
0.026        & 0.56             & 0.45                 & 0.44        & 0.03\\
\hline
0.028        & 0.67             & 0.62                 & 0.49        & 0.06\\
\hline
0.030        & 0.86             & 0.43                 & 0.56        & 0.07\\
\hline   
0.032        & 1.46             & 0.82                 & 0.66        & 0.10\\
\hline     
0.034        & --               & --                   & --          & --\\
\hline
\hspace{0.15in} 0.040 \hspace{0.15in}        &\hspace{0.15in} -- \hspace{0.15in}               
& \hspace{0.15in} -- \hspace{0.15in}         &\hspace{0.15in}  --  \hspace{0.15in}         
&\hspace{0.15in}  -- \hspace{0.15in} 
\end{tabular}
\end{center}
\caption{\label{sphex:tabS} Variation of the exponentiation time-scales with the 
softening parameter $\epsilon_{p}$. Other symbols are as in Table~\ref{sphex:tabN}.
The Ricci and scalar curvatures are positive for the last two values. For 
this reason no exponentiation time-scale is defined}
\end{table}

 In Table~\ref{sphex:tabS} are shown the exponentiation time-scales  calculated
from the values of the Ricci and scalar curvatures averaged over the first crossing time
for  the various simulations that were run.
Since the exponentiation time-scale has an inverse square root dependence on the 
Ricci curvature, it is relatively insensitive to this quantity unless it is
close to zero. The range of softening parameters in which
the transition occurs from negative values of $r_{\bf u}$ --- corresponding to 
an exponentiation time-scale of one dynamical time --- and positive ones is rather 
small. Therefore, although in principle arbitrarily large exponentiation time-scales
can be obtained by fine tuning the softening parameter, in practice a 
softened system appears to have an exponentiation time-scale of the order of a 
dynamical time or none at all. 

From Table~\ref{sphex:tabS} one can also see that, for large softening, 
the exponentiation time-scales associated with the Ricci 
curvature are rather large compared to those
  derived from the average value of the  scalar curvature. This is
because the second term in Eq.~\ref{eq:ru}, which is not present in the formula for the 
scalar curvature, 
is {\em always} positive when significant
softening is present. In turn this is because this term has contributions which contain
the second derivatives in the form
\begin{equation}
\sum - \frac{\partial^{2} W}{\partial q_{i}^{2}} u_{i}^{2}.
\end{equation}
In a spherically symmetric system with positive density decreasing towards the outside,
this expression is positive. It can be fairly large when the softening is large. However
it is still highly fluctuating --- hence the large errors in the exponentiation time-scales. 
This is of course in addition to the last term in~(\ref{eq:ru}) which is also present in the formula for the scalar curvature and is always positive for
systems with positive density.

A completely analogous situation transpires when one increases $N$ with fixed softening and
no short range cutoff. Independent of the softening length or its precise 
functional form, the Ricci
curvature is positive and its average over $N$ remains constant as one varies the particle
number. This is easy to understand since particles which will contribute significantly 
to the local density  at a given point are those within a distance of the order of magnitude
of the softening length $b$. For high $N$ the number of particles in such spheres increases
as $N b^{3}$ while the contribution to the density from each of the particles goes down as
$1/b^{3}$. 

At first sight, the above results may be  taken to mean that the negative Ricci curvature 
is somehow 
caused by the singularity in the Newtonian potential or due to the contribution of nearby
neighbours --- and again the  predicted instability may be a mathematical artifact. 
A closer look however reveals that this is not so.
Near the transition from negative
to positive curvature, the assumptions justifying the validity of the negativity
of the Ricci curvature as an indicator of average instability break down (since
in this case many of the two dimensional curvatures will already be positive and
the variation in their absolute values may be very large). The positivity of the Ricci
curvature in this case only means that the {\em relative} motion of nearby particles 
is regular since the (large) force between them is approximately constant with distance.
These contributions  dominate the value of the Ricci curvature. A similar situation for example
occurs if one encloses the system in  an ``elastic sphere'' where particles near the boundary
are subjected to a harmonic potential. If the spring constant is large, the curvature is dominated
 by the resulting positive contributions from these particles (this experiment was actually conducted
by the author).
 Obviously the nature of the gravitational dynamics in this case is not radically modified.

The transition to positive curvature therefore should not be viewed as indicating 
a sharp switch from a chaotic
to an integrable system but from one where the majority of trajectories
were highly unstable to one where their instability is somewhat less pronounced (how pronounced
can only be determined by calculating the two dimensional curvatures and 
integrating the full set of linearised equations). 
Only in the true
continuum limit (infinite $N$ and fixed softening) is full integrability recovered
(it is interesting to note that the equivalence of the continuum limit to the large-$N$ limit
can only be proven for twice differentiable potentials which are bounded everywhere: 
Braun \& Hepp 1977; Spohn 1980).  
Nevertheless, the fact that the curvature is affected by softening does suggest that the
origin in the instability of the trajectories of gravitational systems is related to their
discrete nature --- this is in line with the fact that smoothing out the force field eliminates
this instability. The effect of softening on the exponentiation time-scale was studied directly
by Suto (1991). He found that the exponentiation time-scale was related to the softening by
$\tau_{e} \sim 20 b^{1/2} \tau_{c}$ (where $b$ is in units of average interparticle distance). 
Thus, while the softening does increase the relaxation time-scale, it does it in a rather moderate
manner. On the other hand, only including the contribution due to nearby neighbours in the
calculation of the curvature yields an exponentiation time-scale that increases as 
$N^{1/3} \tau_{c}$ (Gurzadyan \& Savvidy 1986).
We therefore conclude that while the trajectory instability in gravitational systems 
appears to be related to their discrete nature, what  gives rise to this property
should  be  the contribution to discreteness  noise from  the whole system and not just from 
neighbouring particles. We discuss this further in the concluding section.

\section{The effect of rotation}
\label{sphex:rotsy}

In El-Zant (1996a) it was found that self gravitating sheets starting 
from states of solid rotation initially 
 had a larger  average  Ricci curvature 
 than ones starting from random initial conditions
in velocity space. This property is important since the fact that many flattened 
disk galaxies consist predominantly of stars moving on nearly circular trajectories
near the disk plane suggests that for such systems phase space diffusion time-scales
must be relatively long. In this section we study in more detail the change in the
Ricci curvature as the energy in rotational motion is increased. 
  
Constructing stable equilibrium spherical self  gravitating systems with a wide range of 
ratios of rotational to random motion and having the same density distribution
is not a trivial task (e.g., Palmer 1994), we therefore stick to 
the simple situation of static averages. The systems chosen here are homogeneous and ---
in the absence of rotation --- have isotropic velocity distribution which does not
vary with radius. Solid body rotation is then added to the random motion before all
velocities are rescaled so as to have a total energy of -0.25 in accordance with the
system of units discussed above (Section~\ref{sphex:secn}). 
As we have done before, we calculate the Ricci  curvature for ten different pseudorandom
realizations of the same distribution, calculate the average, and estimate the RMS error (since the scalar curvature does not explicitly depend on the velocities
it remains unchanged when the velocity distribution is changed).

\begin{table}
\begin{center}
\begin{tabular}{c|c|c}
\multicolumn{3}{c}{ }\\
\hline
$T_{rot}/T$ & $\tau_{er}$  &  $\epsilon_{r}$\\ 
\hline
0.00        & 0.30             & 0.07\\              
\hline
0.02        & 0.43             & 0.08 \\                
\hline                 
0.15        & 0.46             & 0.08\\                 
\hline
0.33        & 0.51             & 0.08\\                
\hline
0.55        & 0.61             & 0.08\\                 
\hline   
0.70        & 0.71             & 0.07\\                 
\hline   
0.81        & 0.84             & 0.07\\                
\hline                 
0.90        & 1.02             & 0.06\\                 
\hline
0.95        & 1.18             & 0.05\\                 
\hline
0.97        & 1.26             & 0.04\\                
\hline   
0.98        & 1.32             & 0.04\\                 
\hline   
0.99        & 1.39             & 0.03\\                 
\hline                 
\hspace{0.15in} 0.995 \hspace{0.15in}      & \hspace{0.15in} 1.42 \hspace{0.15in}           
 & \hspace{0.15in} 0.03 \hspace{0.15in}                
\end{tabular}
\end{center}
\caption{\label{sphex:tabR} Variation of the exponentiation time $\tau_{er}$,
averaged over ten (pseudo)random realizations of $N=1000$,
as the fraction of kinetic energy of rotational motion $T_{rot}/T$ is increased.
$\epsilon_{r}$ is an error estimate obtained by using Eq.~(\ref{sphex:ercur})}
\end{table}

Table~\ref{sphex:tabR} shows the variation of  the corresponding exponentiation
time-scales as the energy in rotational motion is increased. As is clear from these 
results, the exponentiation time-scales are significantly increased when the rotational
motion is increased. Looking at the different terms on the right hand side of 
Equation~(\ref{eq:ru}), it 
is easy to see that only the first two are directly dependent on the velocities
and may therefore be affected by the reordering  of random motion into rotational motion. 
Of these two terms the first --- as discussed above --- is small for most equilibrium velocity
distributions. It is even smaller for rotating systems
since the 
sum of the scalar product of the particles' velocities and the forces acting on them which
this term represents is near zero.  
The second term consists
of two parts. One involves the quantities in~(\ref{eq:rut2}) and
represents the second derivatives of the potential with respect
to coordinates of the same particle multiplied by the velocities of that particle. This
term is also relatively small since again the gradient of the force in a corresponding
smoothed out system is normal to the velocities (albeit not as small as the first term 
since the components of the derivatives of the force have  larger fluctuations than
the components of the force).  
The other 
part of the second term (involving the quantities in~(\ref{eq:rut1}))  consists of
the derivatives of the force at one particle's position with respect to another's projected
on the velocities of the two particles --- in other words it measures  
correlations in velocities and positions between the trajectories of particles in the systems. 
This term is small and fluctuating when the velocity field is random but
is much larger and has positive sign when the kinetic energy is in the form of ordered
rotational motion. 

The fact that the exponentiation time-scale is significantly larger when 
ordered motion is present confirms that, except when effects arising from the
non-compactness of the phase space are important (e.g., escape of particles),
higher thermodynamic entropies appear to be related to higher values for the 
dynamical entropy. Thus confirming that $N$-body systems will, in general, evolve
towards higher dynamical entropy states as was found to be the case in El-Zant \&
Gurzadyan (1997). 

If directly interpreted to mean mixing on an exponential rate as is expected in the
standard case of gravitational systems with negative two dimensional curvatures, 
the derived time-scales predict that a rotating system will evolve on a time-scale 
of about 45 crossing times (El-Zant 1996a). This is still compatible with the age
of average disk galaxies at about 10 kpc, and it is possible that for realistic
density and velocity distributions the predicted evolution time-scale may be still larger.
In addition, enough  two dimensional curvatures may be positive so as to restrict 
the motion.
 Therefore, as in the case of softened systems, the variation of the Ricci curvature
exponentiation times with rotation should only be interpreted as representing a trend. 
However, as we shall see in the next section, it appears that the relation between
the exponentiation time-scales and the macroscopic evolution of gravitational systems
may not be so direct.

\section{Conclusions and possible interpretation}     

In this paper we continued our investigation of  the behaviour of the Ricci and scalar
curvatures of the configuration manifolds of $N$-body gravitational systems. 
These relate the geometry of the phase space to the stability properties of
trajectories on it.
It was
found that,  for spherical systems with isotropic velocities,
 the inferred exponentiation
time-scale is rather short (less than a crossing time) and did not depend
on the particle number  when the softening length decreased  as $1/N$ and a short range cutoff in the
potential was introduced. 
The exponentiation time-scale however was found to
be affected by the presence of ordered rotational motion or when the softening
was increased (while keeping the particle number fixed). 
In the first case it was found to increase significantly while
in the second case it could even become undefined because the curvature became
positive. A similar process takes place if the softening radius is fixed and 
the number of particles is increased (if no short range cutoff is introduced).

The exponentiation time-scale being so short and not varying with particle 
number means that it is difficult to uncover its significance.
The main problem is that these properties apparently contradict the intuitive
idea that particles in large systems should move essentially unperturbed in the
mean field potential. In spherical potentials this happens to mean that they
all lie on regular trajectories. It then should follow that the divergence between
nearby trajectories is, on average, linear and not exponential. 

One may like to
relate the exponential instability to the process of achieving {\em dynamical} 
equilibrium,
the time-scale of which does not vary with particle number.
(e.g., Kandrup 1989; El-Zant 1996c). 
While this could be the case, one expects that even
completely smooth spherical systems achieve such an equilibrium. 
The exponential instability on the other hand 
appears to be inherently related to the discreteness of $N$-body gravitational systems.
At the same time this does not imply that it is mainly a result 
of  close encounters as was explained
near the end of Section~\ref{soft}. 
In the light of that discussion, and looking at the relative importance of the  terms of the formula
for the Ricci curvature (see Section~\ref{sphex:secn}), one comes  to the conclusion that, for spherical systems 
with isotropic velocity distribution and in virial equilibrium, the negative curvature is
related to the fact that $N$-body systems have a large mean field force (because the
interaction is long range) and at the same time have locally peaked  density
distribution (because of their being composed of particles). 
Thus we may expect that the cause of the exponential
instability is the discreteness effects due to the long range full $N$-body interaction.

To see how long range gravitational
interaction can trigger chaotic behaviour, we follow Chirikov (1979) and divide 
the  Hamiltonian
 of the system under consideration
into an unperturbed part $H_{0}$ and an non-integrable time dependent perturbation $V$.  In terms of action
angle variables this reads 
\begin{equation}
H({\bf J}, {\bf \Theta}, t)= H_{0}({\bf J}) + V ({\bf J}, {\bf \Theta}, t).
\end{equation} 
For our purposes $H_{0}$ will be the smooth spherically symmetric potential and $V$ would the
perturbation arising from discreteness effects: $V= V_{nbody} - V_{0}$.
If we assume  that $V$ is periodic in time with phase $\tau = \Omega t + \tau_{0}$,
 there will be a whole set of resonance conditions given by
\begin{equation}
m_{i} \omega_{i}({\bf J}_{i}) + n \Omega =0,
\end{equation}
where $\omega_{i}$ ($i=1,3$) are the natural frequencies of oscillation of the action variables 
of the motion in the background potential $V_{0}$. Around the neighbourhood of these
 resonances a stochastic layer on which chaotic motion can take place forms 
(e.g., Lichtenberg \& Lieberman 1983 (LL)).
If $\Omega >> \omega_{i}$ however,
the  resonance conditions are satisfied only for very large $m/n$ and no lower order resonances occur.
In this case the effect of these resonances is limited because the stochastic layer around them 
is small (LL; Meiss 1987).
The vast majority of non-resonant trajectories will therefore remain stable.
In the six dimensional phase space this will mean  the perturbation causing trajectories to move
between KAM tori. 
In that case the original idea of
Chandrasekhar (1942) is justified: he considered discreteness effects  due to short lived
encounters with nearest neighbours where resonance effects are negligible 
and where there is a clear separation of time-scales justifying
the assumption of independence. 
However it is now thought that weak distant 
encounters dominate two body interactions 
in $N$-body systems and that distant encounters are more important 
for larger systems (BT).
Thus, although the strength of perturbations due to discreteness 
decreases as $\sim 1/\sqrt{N}$,
the density of resonances (per action per particle) increases as $\sim N$ and
is increasingly dominated by more effective terms.
This might explain the observed  persistence of chaotic behaviour
for large $N$. For, according to LL, this issue reduces 
to ``the question of whether the density of important resonances, as projected at 
a single action, increases faster than the width of the resonances decreases.
If this happens then we would expect resonance overlap and strongly chaotic 
motion to occur for $N$ degrees of freedom as $N \rightarrow \infty$''.

It is important to note here that although the main cause of the chaotic behaviour may be 
distant interactions between particles in  a $N$-body system, this does not mean that the exponential instability
will not be affected by short range encounters. For, even if one considers these as additional
high frequency  noise added to the system,
according to Pfenniger (1986) such perturbations completely change a chaotic system's trajectory 
on relatively short time-scales. One therefore expects high frequrency discreteness noise to be an 
additional source of instability which will affect the exponentiation time-scale. This would explain 
why, even if the instability is mainly caused by  distant encounters, the exponentiation time-scale 
increases when the force  softening is increased this suppressing the high frequency noise.

It is clear that as $N$ increases, quantities such as the energy and angular momenta of
individual particles will be better conserved.
 This however does not imply that the decorrelation time-scale defined
by the exponential divergence should become smaller. 
For example a perturbed 
pendulum can display highly chaotic oscillations which decorrelate  very fast while changes in the 
amplitudes of these oscillations are much slower. In this type of situation 
one can average over the fast phase variable which may (because of the short decorrelation time) be considered
as random. The  evolution of the actions can then be regarded as a diffusion process. In fact this approximation
is only valid when the dynamics is {\em strongly  chaotic} so that chaos
occurs for the vast majority of initial conditions (Chirikov 1979; LL; Shlesinger et al. 1993).
This is of course what appears to be the case for gravitational $N$-body systems. 

In the case of gravitational systems, it is actually not that surprising that the exponentiation time-scales are very
different from the diffusion time-scales of the action variables. For in the standard case of a system with negative two dimensional
curvatures, the exponential instability will guaranty that the system visit all regions of the available 
phase space (Anosov 1967).
 In the case of a $N$-body system this will be the whole subspace defined by the conservation of
total energy and momentum. Since in a Hamiltonian system the phase density is conserved along the motion, this will imply that
the phase space ($N$ particle) distribution function will become constant when averaged over   progressively smaller volumes in that space ---
that is the system becomes  more and more  likely to be found in any of its microscopic states.
For an open gravitational system this clearly cannot be the case since,
instead of evolving towards a definite
thermodynamics equilibrium, this type of system continually evolves towards more and more concentrated 
configurations  when it divides into a contracting core section surrounded by an expanding halo.
In this type of evolution, the Poincar\'e recurrence theorem is not valid and the system need not, even
in principle, return to less inhomogeneous states. Instead it continually moves into new areas of the 
phase space characterised by larger entropy. In this case, if the evolution time was directly related to
the exponentiation time-scale, then there would be no chance for the distribution function to be constant
on any region of the phase space. 
Therefore, no type of equilibrium would be possible and gravitational
systems would disintegrate in a few dynamical times!
 What prevents this from happening of course is that some states are long lived because 
they are stable dynamical equilibria. 

Indeed, it was found that, for closed systems which did have a definite {\em thermodynamic} equilibrium
state, this state was reached on a time-scale comparable to the exponentiation time if no dynamical equilibrium
existed between 
the initial state and the final equilibrium (this happened even when the virial ratio remained
nearly constant during the evolution thus ruling out violent relaxation). On the  other hand, if there existed
intermediate {\em dynamical} equilibrium states between the final equilibrium and the initial configuration,
the relaxation to the final equilibrium state took much longer (El-Zant 1997).
It may be therefore that the exponential instability acts as to smooth out the distribution function
on the subspaces where these equilibria are defined while at the same time driving the evolution towards higher
entropy states. This means that although the instability is locally normal to the phase space motion, the global
stretch and fold of that space is such that the decorrelation is faster along the motion --- 
justifying the diffusion description for the evolution of the action variables. For larger $N$, the dimension of the phase
space grows and there are more states (corresponding to a given dynamical equilibrium) to be covered while
the exponentiation time-scale  is constant. 
Thus, one expects the diffusion rate of the action variables to be larger as $N$ increases.

More precisely, it can  be shown using 
symbolic dynamics (Alexseev \& Yakobson 1981;
McCauley 1993; Chirikov 1994) 
that for hyperbolic systems (a class which includes systems with negative two dimensional curvatures),
with well defined final states,
 the time-scale for a system to achieve such a state with coarse grained volume resolution 
 $\mu$ is
\begin{equation}
\tau_{r}=\frac{\ln (1/\mu)}{h},
\end{equation}
where $h \sim \langle \sqrt{-2 r_{\bf u}} \rangle$ is the Kolmogorov entropy
in the case  when $r_{\bf u}$ is dominated by 
the largest two dimensional curvatures. 
(The above  relation is exact and  follows from equation (14.3) of 
Alexseev \&  Yakobson when $h$ is constant along the $N$-body trajectory, 
otherwise it is valid in an averaged asymptotic sense). 
Fixing the linear resolution $d$
and using Eq~\ref{dyn} one finds
\begin{equation}
\tau_{r}= 2 \sqrt{3} \ln (1/d) W \sqrt{N}  \tau_{e}  \sim \sqrt{N}.
\end{equation}  
Therefore, in  systems for which  definite statistical equilibria exist, the time-scale
to achieve such equilibria does indeed in general increase with $N$ and appears to
scale as the inverse of the strength of the perturbation due to discreteness. This is the
case even though the exponetiation (or Liapunov)
time-scale as derived  from the value of the Ricci curvature 
(or as directly inferred from the divergence of 
$N$-body trajectories) does not vary much with $N$. 
This is essentially because, unlike the case of low dimensional systems, 
the Kolmogorov
entropy is dominated by the maximal exponent and volume resolution becomes 
very different from linear resolution.  
 Thus, for large-$N$ systems,
the Liapunov time-scale does not necessarilly coincide with the mixing time-scale.  
A more detailed
discussion of the mathematical origin of this important effect, 
which in retrospect has been a major cause of the confusion
regarding  the meaning of the Liapunov time-scale for $N$-body systems, 
will be given elsewhere.

Because of the exponential instability, the  (6-d) phase space  
trajectories of particles quickly decorrelate so that the probability of finding a particle
in a given state is independent of the state of the other particles. This being so even if the
motion is affected by the discrete nature of the system.
For times longer than the decorrelation time $\tau_{e}$ 
the $N$ particle distribution function can  then be given in terms of the individual one particle
distribution function  by
\begin{equation}
f^{N}=f({\bf x}_{1}, \dot{\bf x}_{1},t) f({\bf x}_{2}, \dot{\bf x}_{2},t) ...
f({\bf x}_{N}, \dot{\bf x}_{N},t).
\end{equation}
This is of course the requirement that a system be completely described by the collisionless
Boltzmann equation (CBE). 
Moreover,
since steady state solutions of the CBE will only depend on the action variables (Jean's theorem),
these solutions will be equivalent to those produced by the mean field dynamics ---  for times long compared
to the exponentiation time-scale but short compared to the diffusion rate of the action variables. 
In this context the continuum (collisionless) approximation may then be valid in an {\em averaged} sense: the exact trajectories
in smoothed background potentials would not be valid but the time averaged {\em orbits} 
would be correct as long
as the time-scales considered are small compared to the diffusion times of the action variables.

Although for many purposes the situation described above 
 is similar to that of standard  galaxy dynamics, it differs
in one  important respect: {\em the underlying motion is  intrinsically chaotic
and cannot be expressed as a linear superposition of regular motion and independent binary
encounters}. 
For this reason
 the diffusion time of the action variables need 
not be accurately represented by the two body relaxation time.
Moreover, it is now well known that chaotic trajectories 
can repond to external perturbations in a manner that is different 
from that of regular trajectories (Pfenniger 1986; Kandrup 1994; Merritt \& Valluri 1996; El-Zant 1996c). 
Therefore, the response of the trajectories to additional perturbations
(e.g., high frequency discreteness noise or global asymmetries in the background 
potential) is not necessarilly identical to that of the trajectories in the 
smoothed out potential --- which for spherical systems happen to be all regular.

To sum up. In this section it has been suggested that resonances between the orbital frequencies
of particles and forcing caused by the dicreet nature of the global potential give 
rise to chaotic trajectories which decorrelate over a time-scale of the order of a dynamical
time. 
Nevertheless some quantities  may decorrelate on the much
larger time-scale. In the case of systems of having a well defined final equilibrium state 
the maximum such time-scale is
that needed to reach a statistical equilibrium when such a state exists which scales as
$N^{1/2}$.

The above ideas can be tested in more than one way.
As mentioned in the introduction, it is well documented that $N$-body
trajectories decorrelate over a dynamical time or less. 
It may be useful however to also
examine the stability of {\em individual} particle  trajectories in $N$-body simulations --- preferably by methods
that do not require linearisation of the equations of motion (e.g., Laskar's frequency analysis).
 To find out the dominant range of encounters causing the exponential divergence
in $N$-body systems, it may  be possible to calculate the frequency spectrum of perturbations a 
$N$-body particle is subjected to and examine the effect of its different regions on individual
trajectories in smoothed out potentials. Alternatively, one may 
 want to bin the contributions  to the discreteness noise 
${\bf F}={\bf F}_{smooth}-{\bf F}_{nbody}$ acting along 
a particle's trajectory into impact parameter ranges and again examine their effect on the stability 
of test particles' trajectories  in the smoothed out density distribution. 
The time-scales over which the physical characteristics of a given system changes can be studied
directly  by examining the relaxation of the particles' integrals of motion in the smoothed out potential.
This can be done either by computing the diffusion
coefficients of these variables for particles in $N$-body simulations 
or by studying the macroscopic relaxation of numerically simulated systems.
The latter approach is  most effective
 for closed systems where definite thermal equilibria are well defined
(El-Zant 1996b). One would then look at the time-scale for attaining isothermal
equilibrium and the time-scale of relaxation of initial anisotropic
velocity distributions (El-Zant 1997) and how these time-scales vary
with $N$ (El-Zant \& Goodwin 1997). In fact, the   existence of a well defined final 
state that ceases to evolve also means that all the aforementioned tests are easier to conduct 
for  closed  systems.

\section*{\em Acknowledgements}
I would like to thank  Vahe Gurzadyan for many interesting discussions on
the subject of this paper and for commenting on the manuscript.
It is also a pleasure to thank John Papaloizou and Peter Thomas for
helpful discussions. Thanks must also go to Sverre Aarseth for providing a
copy of his NBODY2 code and Simon Goodwin for helping with its use.

\newpage
\section*{Figure caption}

\begin{flushleft}
\begin{figure}
\caption{\label{softcur}
 Evolution of the Ricci curvature for $N=1000$ systems with different  values for the
softening parameter $b$ which  takes the values of (from top to bottom)
 $4.0 \times 10^{-3}$, $1.6 \times 10^{-2}$, 
$2.8 \times 10^{-2}$ and $4 \times 10^{-2}$}
\end{figure}
\end{flushleft}

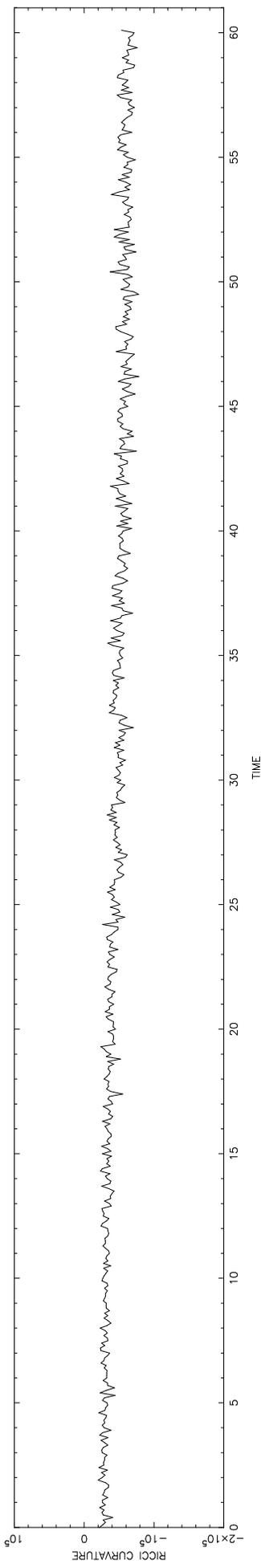
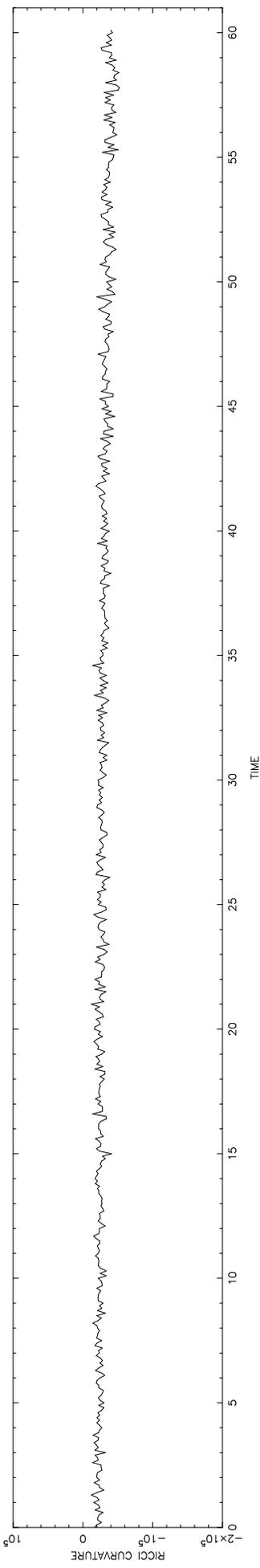
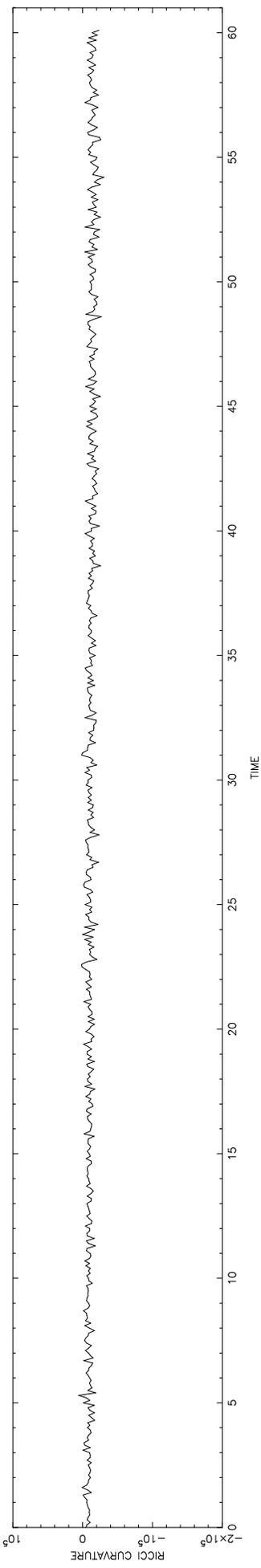
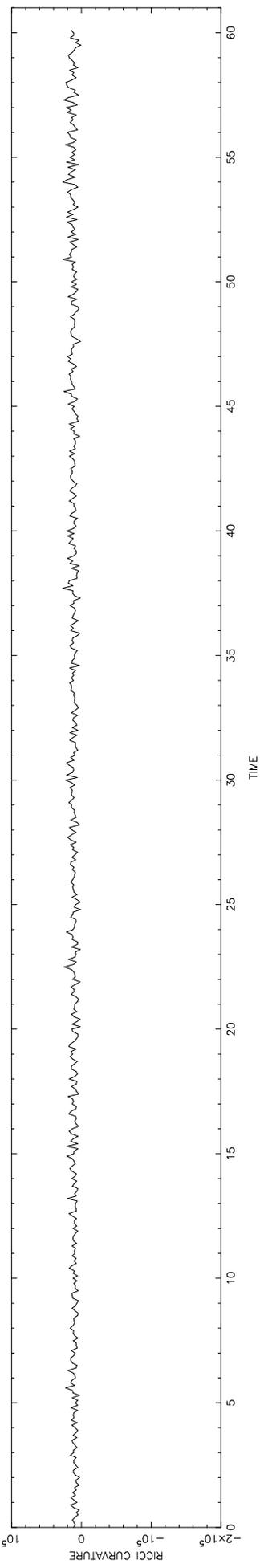
\end{document}